\documentstyle[11pt,paspconf,psfig]{article}

\begin{document}

\title{Dark Matter in Late-type Dwarf Galaxies}

\author{Rob Swaters}
\affil{Kapteyn Institute, 9700 AV Groningen, The Netherlands}


\section{Introduction}

Previous studies of dwarf irregular galaxies have mostly found that
these systems have slowly rising rotation curves and that they are
dominated by dark matter, even well within the optical disk (e.g.,
Carignan \& Beaulieu 1989, Broeils 1992, C\^ot\'e 1995). However,
based on new observations and a different technique to derive rotation
curves, we find that the rotation curves may be steeper and that dark
matter may be important in the outer parts only.\vspace{-2mm}

\section{Sample and observations}
\vspace{-1mm}
From the UGC catalog, we have selected all galaxies with Hubble types
Sdm or later, north of $\delta = 20$ and with HI flux densities larger
than 200 mJy.  This resulted in a sample of 112 dwarf galaxies, 75 of
which have been observed in HI with the Westerbork Synthesis Radio
Telescope as part of the WHISP survey (Kamphuis et al.\ 1996), and in
the optical $R$-band with the Isaac Newton Telescope at La
Palma (Swaters \& Balcells, in preparation).\vspace{-2mm}


\section{Mass modeling}
\vspace{-1mm}
Rotation curves have been derived for 44 dwarf galaxies that have high
enough signal-to-noise ratios, that are sufficiently resolved, and
that show clear signs of rotation.  Mass models have been fitted to
these rotation curves, assuming ``maximum disk'', i.e.\ scaling the
stellar contribution to the rotation to fit most of the inner part of
the rotation curve.  For the dark matter distribution an isothermal
halo with core has been assumed.

Initially, the rotation curves were derived by fitting a tilted ring
model to the velocity fields.  Fig.\ 1 shows the rotation curve for
one example: UGC~7559. The mass model based on this rotation curve
(see Fig.\ 1) shows that the dark matter dominates at almost all
radii.  However, as can be seen from the fit, the maximum disk scaling
hinges upon the very inner points, and only a small change in their
values can have a strong effect on the derived mass model.

\begin{figure}[ht]
\psfig{file=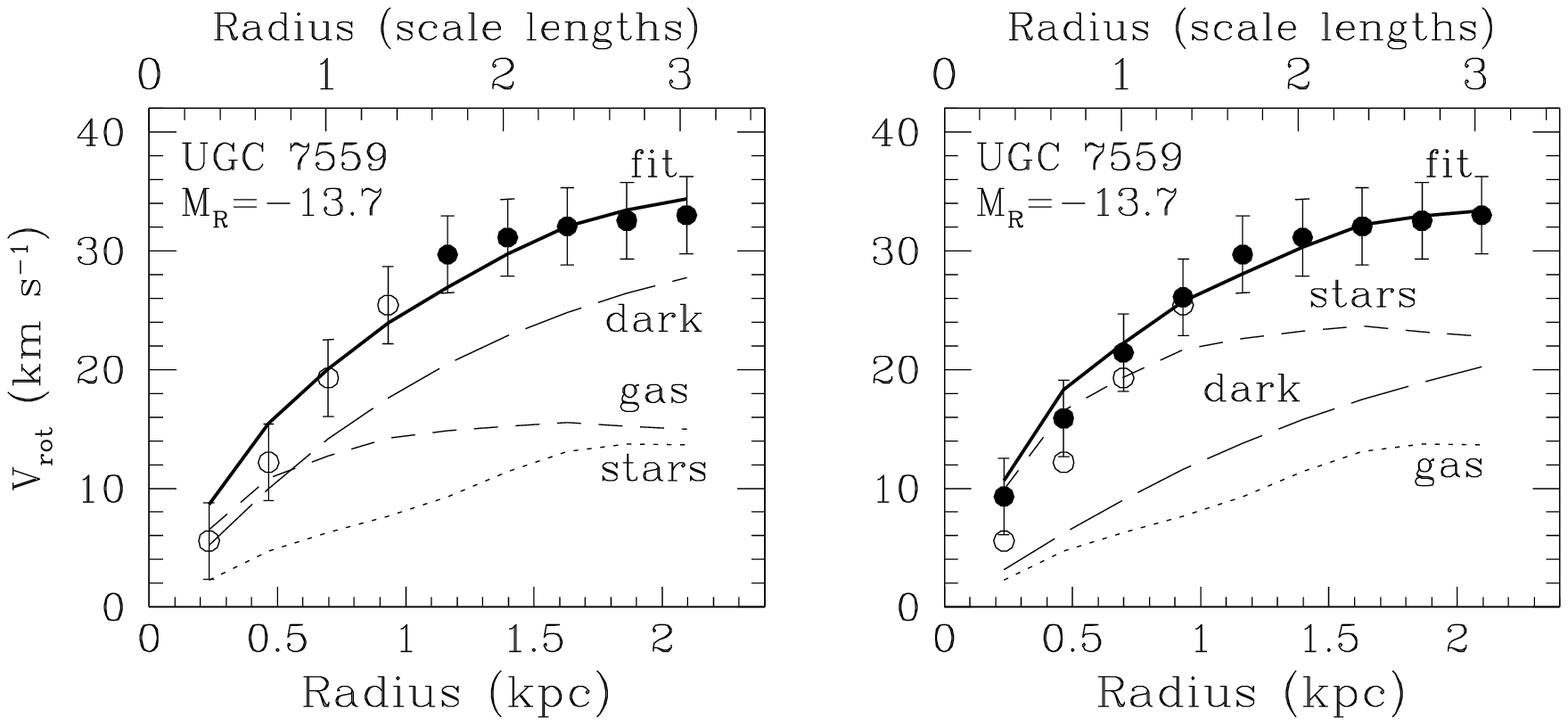,height=6cm}
\vspace{-0.2cm}
\hbox to\hsize{
\vtop{\hsize=6.2cm
Figure 1. Maximum disk model for UGC~7559 for the tilted ring rotation
curve}
\hfill
\vtop{\hsize=6.2cm
Figure 2. Maximum disk model for UGC~7559 for the modelled rotation
curve}
}
\vspace{-0.4cm}
\end{figure}

To estimate the uncertainties in the inner points, due to the combined
effects of beam smearing and the distribution of HI, we constructed
model data cubes for these dwarfs, using as input the observed HI
distribution, the derived rotation curve and a velocity dispersion of
the gas of 8 km s$^{-1}$. The output model was compared to the
observations, and the rotation velocities were scaled up to find the
steepest rotation curve that is still compatible with the data.  Fig.\ 
2 shows the mass models for the same galaxy, based on the modelled
rotation curve.  Even though the changes in the rotation curve are
only a few km s$^{-1}$, the mass model has changed dramatically.
Rather than being dominated by dark matter, the stellar distribution
dominates the mass within the optical radius for maximum disk, and the
dark mass starts to take over only beyond three disk scale lengths.
This is true not only for UGC~7559, but for all galaxies in this
sample. However, the maximum disk models require high stellar $M/L$
values for some of the dwarf galaxies.  For the modelled rotation
curves, a third of the $M/L$ values are larger than 5.

Within the uncertainties in the rotation curve in the inner parts,
both solutions, dominant dark halo or dominant stellar mass within the
optical radius, can be accomodated.  More detailed modelling and
higher resolution observations are necessary to pin down the exact
shape of the rotation curve in the inner parts.  H$\alpha$ rotation
curves for a few dwarf galaxies show that the inner rotation
curves indeed are steeper than those derived using titled ring fits
(Carignan 1985, Blais-Ouellette, this volume).

The dwarf galaxies in this sample show that the canonical
picture of dwarfs being dominated by dark matter within the optical
disk may not be correct.  The rotation curves for dwarfs may be
steeper and may be dominated by the stellar contribution within the
optical radius.  Outside the optical disk, dark matter dominates.

\end{document}